\documentclass[%
 reprint,onecolumn
 amsmath,amssymb,
 aps,
]{revtex4-2}
\usepackage[normalem]{ulem}
\usepackage{amsmath}
\usepackage{float}
\usepackage{xcolor}
\usepackage{graphicx}
\usepackage{dcolumn}
\usepackage{bm}

\newcommand{\blu}[1]{\textcolor{blue}{#1}}

\newcommand{\gs}[1]{\textcolor{blue}{GS: \textbf{\textit{#1}}}}

\begin{document}

\preprint{APS/123-QED}

\title{On the Dynamics of Liquids in the Large-Dimensional Limit}

\author{Chen Liu}
\affiliation{Laboratoire de Physique de l’École normale supérieure, ENS, Université PSL, CNRS, Sorbonne Université, Université de Paris F-75005 Paris, France}
\affiliation{Department of Chemistry, Columbia University, New York,
New York 10027, USA}

\author{Giulio Biroli}
\affiliation{Laboratoire de Physique de l’École normale supérieure, ENS, Université PSL, CNRS, Sorbonne Université, Université de Paris F-75005 Paris, France}

\author{David R. Reichman}
\affiliation{Department of Chemistry, Columbia University, New York,
New York 10027, USA}

\author{Grzegorz Szamel}
\affiliation{Department of Chemistry, 
Colorado State University, Fort Collins, CO 80525,
USA}


\date{\today}

\begin{abstract}
In this work we analytically derive the exact closed dynamical equations for a liquid with short-ranged interactions in large spatial dimensions using the same statistical mechanics tools employed to analyze Brownian motion. 
Our derivation greatly simplifies the original path-integral-based route to these equations and provides new insight into the physical features associated with high-dimensional liquids and glass formation.  Most importantly, our construction provides a facile route to the exact dynamical analysis of important related dynamical problems, as well as a means to devise cluster generalizations of the exact solution in infinite dimensions. This latter fact opens the door to the construction of increasingly accurate theories of vitrification in three-dimensional liquids.
\end{abstract}

\maketitle

{\it Introduction} -- The motion of interacting particles in a liquid is so complex that a complete, microscopic description of liquid state dynamics generally requires {\em in silico} experiments that directly integrate the underlying Newtonian or Brownian equations of motion one particle at a time.  For supercooled liquids, however, these simulations are impossible to perform close to the glass transition, as the drastic slowing of dynamics precludes the possibility of modern-day processors from describing long-time relaxation via this painstaking technique. Instead, one generally relies on approximate microscopic and coarse-grained theories to gain an understanding of the long time dynamical behavior of complex processes such as vitrification.

The main difficulty in developing such theories is that 
glassy slowing down is a strongly interacting problem which eludes perturbative treatments \cite{BBRMP}. One important aspect of the glass transition is that the dramatic growth of the relaxation time is accompanied by a very modest growth of the length scale $l_\text{coop}$ characterizing the spatial extent over which cooperative motion takes place~\cite{BBRMP,karmakar2015length}.
A theory able to accurately describe dynamics over such length scale $l_\text{coop}$ would therefore provide a complete description of the phenomenon. In the case of strongly correlated electrons, a problem that shares similar technical challenges, following the path paved by this intuition has paid off handsomely via the creation of a dynamical mean-field theory (DMFT) able to describe the physics at the shortest scale~\cite{KRMP}, and then cluster extensions able to capture non-perturbative physics below and at the scale $l_\text{coop}$~\cite{JarrellRMP,cellDMFT}. Developing an analogous approach for glassy liquids is of tremendous importance.  In this work we focus on the very first step, which is development of DMFT for liquids. The recent exact solution of glassy liquids in the limit of infinite dimensions was a complete breakthrough in this respect. Using two independent routes, a {\it super-symmetric} path-integral treatment and {\it an approximate cavity approach}, DMFT for the dynamics of interacting particle systems was derived in the $d\rightarrow \infty$ limit ~\cite{kurchan2016statics,dmft,agoritsas2019out} \footnote{The cavity derivation of Ref. \cite{agoritsas2019out} was inspired by G. Szamel,
Simple Theory for the Dynamics of Mean-Field-Like Models of Glass-Forming Fluids,
Phys. Rev. Lett. \textbf{119}, 155502 (2017).}. These two tools, however, cannot easily be generalized to develop cluster methods since the former is somewhat cumbersome whereas the latter is based on some approximations whose validity is unclear. The aim of this work is to present a general approach to obtain a liquid-state DMFT that is direct, versatile and physically transparent, hence suitable to be generalized to more complex cases and in particular to cluster methods.  

As a remarkable byproduct, our approach allows us to bridge the gap between the theoretical methods behind the Mode-Coupling Theory (MCT) of the glass transition~\cite{reichMCT,gotze2008complex} and the techniques at the basis of the Random First Order Transition (RFOT) theory~\cite{KTW,WolReview,parisi2020theory}. 
Initially it was believed that MCT is exact in the limit of infinite spatial dimensions~\cite{KW}.  However, a decade ago it was demonstrated that MCT is actually increasingly {\em less} accurate as the spatial dimension increases~\cite{Kuni,Schilling}.  Such behavior is unexpected for a mean-field theory, and this failure of MCT temporarily clouded the connection between statics and dynamics that lies at the heart of foundational theories of the glass transition such as the Random First-Order Theory (RFOT)~\cite{parisi2020theory}.
As mentioned above, the original derivation 
of the exact infinite dimensional dynamical theory is highly technical and makes use of complex path-integral techniques, which are very different from the projection operators that are standard in statistical mechanics and are used to derive MCT, and thus does not establish a direct connection.  
In this work, by properly identifying the {\it correct tagged degree of freedom}, which allows to treat the rest of the system as a self-consistent thermal bath, we derive DMFT by the projection operator method. We thus unify seemingly disparate routes to dynamics in $d=\infty$ and show how to modify the MCT derivation to obtain the correct infinite dimensional limit.  This unification enables the exact, physically clear description of the large dimensional dynamical behavior of Newtonian and Brownian fluids, opens a simple route to the exact solution of distinct models of slow dynamics (e.g. the Lorentz gas), and sets the stage for the extension of the dynamical mean-field concept to lower space dimensions via the introduction of a ``cluster" dynamical mean-field approach.  We describe all of these facets in this work.  

Our derivation applies to an equilibrium system composed of a large number $N$ of identical particles interacting through a pairwise, short-ranged potential $v(r)$ with the dynamics
\begin{equation}\label{eq:1}
    m\Ddot{\bm{R}}_i + \zeta \dot{\bm{R}}_i = -\sum_{j(\neq i)} \bm{\nabla}v(\bm{R}_{ij})+\bm{\xi}_i,
\end{equation}
where $\bm{R}_{ij}\hat{=}\bm{R}_i-\bm{R}_j$, and the thermal noise satisfies $\langle\xi_\mu(t)\xi_\nu(s)\rangle_{\xi}=2T\zeta\delta_{\mu\nu}\delta(t-s)$. The particles are labeled by $i=0,1,2,\ldots,N$ and the Euclidean components by $\mu,\nu\in\{1,2,\ldots,d\}$. In order to obtain a well-defined large-$d$ limit one has to adopt the following scaling \cite{kurchan2016statics} (see Supplemental Information (SI) for more details): the interaction potential depends on the dimension as $v(r)=\overline{v}(d(r/\ell-1))$, where $\ell$ (the  interaction range) is set to $1$, and $\overline{v}$ is a function independent of $d$. Thus the n-th derivative $v^{(n)}$ is of order $\mathcal{O}(d^n)$. Furthermore, to describe correlated dynamics, which determines long-time transport properties and transient localization upon approaching the glass transition, it is enough to focus on the evolution of the mean squared displacement of individual particles on a scale $1/{d}$, i.e. $\bm{u}_i^2\sim d^{-1}$ with $\bm{u}_i\hat{=}\bm{R}_i(t)-\bm{R}_i(0)$, and accordingly on a scale $1/d$ for a given component, i.e. $u_{i,\mu}\sim d^{-1}$. This scaling of a given {\em component} is crucial, as we will see below.  For consistency, we must also take the following scaling relations  $\zeta\sim d^2$, $m \sim d^{2}$, $\xi_{i,\mu} \sim d$ for all $i,\,\mu$ \cite{dmft}.

We now present our general derivation of the DMFT equations. First, we discuss in detail the key steps of the projection operator-based analysis of the Newtonian case, where the friction term and the noise are absent. We then show how such procedure can be generalized to the full form of Eq.\ref{eq:1}, discuss an alternative derivation based on the cavity method, and relate the two methods. 
As we have already stressed, the key conceptual starting point is to identify the correct tagged  degree of freedom. By using the large-$d$ scaling above, one finds that for times $t \sim \mathcal{O}(d^0)$ the force between a pair of particles $i$ and $j$ is only non negligible along the initial relative direction, i.e. $\bm{\nabla}v(\bm{R}_{ij})\approx \hat{\bm{R}}_{ij}(0)v'(R_{ij})$. This allows us to write the equation of motion for the displacement of particle $i$ along direction $\alpha$ as 
\begin{equation}\label{eq:2}
    m\Ddot{u}_{i,\alpha} + \zeta \dot{u}_{i,\alpha}= -\sum_{j(\neq i)} \hat{{R}}_{ij,\alpha}(0)v'(R_{ij})+{\xi}_{i,\alpha} .
\end{equation}
where all terms are of order $\mathcal{O}(d)$ and hence lead to a well defined equation in $d\rightarrow\infty$ (details in SI). Remarkably, the form of the interaction term is strongly reminiscent of that of mean-field spin glasses, where the role of the disordered quenched magnetic coupling $J_{ij}$ is now played by the $\alpha$ component of the initial distances $\hat{{R}}_{ij,\alpha}(0)$. This suggests that the correct  degree of freedom to develop DMFT is the component of the displacement of a tagged particle (say particle $0$) along a fixed direction $\alpha$ uncorrelated with the 
interparticle directions  $\hat{\mathbf{R}}_{ij}(0)$.
The main physical requirement is that this degree of freedom must act as a small perturbation for the rest of the system, and that this perturbation can be taken into account using linear response theory. This naturally leads to a feedback from the rest of the system which is akin to a thermal bath. 
The theoretical frameworks we develop below show that this is indeed the case for the displacement of particle $i$ along direction $\alpha$, and that other choices of the tagged (or "cavity") variable, such as the full displacement vector of a given particle, do not lead to a closed self-consistent dynamical process. 

The first method we employ is one that has been used in the past to derive the exact Langevin equation for a heavy particle in a bath of light particles. 
Following Mazur and Oppenheim~\cite{MO}, we derive the \textbf{exact} (in any dimension) Langevin equation
\begin{eqnarray}\label{eq:3}
&&\dot{p}_{0,\alpha}(t) = F_{\alpha}^{\dagger}(t) \nonumber \\
&& + \int_{0}^{t}d\tau e^{iL(t-\tau)}(\nabla_{p_{0,\alpha}}-\frac{p_{0,\alpha}}{mk_{b}T})\langle F_{\alpha}F_{\alpha}^{\dagger}(\tau)\rangle_{0},
\end{eqnarray}
where $L$ is the Liouville operator that encodes the Newtonian dynamics, namely $iLA=\{A,H\}$. We call $F_{\alpha}^{\dagger}(t)=e^{i(1-\mathcal{P})Lt}F_{0,\alpha}(0)$ 
($F_\alpha\equiv F_\alpha^\dagger(0)\equiv F_{0,\alpha}(0)$) 
the fluctuating (random) force, where $\mathcal{P}B=\langle B \rangle_{0}$ and the ensemble average is taken 
with respect to the Hamiltonian $H_{0}$ of the system with the tagged particle coordinate along the $\alpha$ direction frozen, defined by $H=H_{0}+\frac{p_{0,\alpha}^{2}}{2m}$, where $H$ is 
the full Hamiltonian.
Note that the gradient with respect to $p_{0,\alpha}$ of the force-force correlation is in principle not zero since $L$ depends on $p_{0,\alpha}$. 
All details can be found in the SI. 

This exact Langevin equation would appear to suffer from an issue typical of projection operator derivations, namely that the difficult to analyze and implement projected dynamics defines the evolution of the fluctuating force and also appears in the memory kernel. In the large $d$ limit this difficulty can be overcome. First, we define the evolution operator $L_0$ that corresponds to the motion with the tagged particle blocked along the $\alpha$ direction, $iL_0 A=\{A,H_0\}$. We use a tilde to denote the dynamics defined by $iL_0$, \textit{i.e.} $e^{iL_0 t} A = \tilde{A}(t)$. In particular, the displacement variable of the tagged particle relative to that of the $j$-th particle when the former is blocked along the $\alpha$ direction is $\tilde{\bm{R}}_{0j}(t) = \bm{R}_{0j}(0)+\tilde{\bm{u}}_{0,\alpha}^{\perp}(t)-\tilde{\bm{u}}_j(t)$ in a self-explanatory notation. We also distinguish the force with the $\alpha$ direction blocked, namely $\tilde{F}_{0,\alpha}(t)=e^{iL_0t}F_{0,\alpha}(0)=-\sum_{j(\neq 0)} \nabla_{\alpha}v(\tilde{\bm{R}}_{0j})$ from the unconstrained force along the $\alpha$ direction, namely $F_{0,\alpha}=-\sum_{j(\neq 0)} \nabla_{\alpha}v(\bm{R}_{0j})$.
Then, we isolate the part of the projected evolution operator that corresponds to the motion with the tagged particle blocked along the $\alpha$ direction, $$(1-\mathcal{P})L=L_0+(1-\mathcal{P})\left[\frac{p_{0,\alpha}}{m}\nabla_{0,\alpha}+F_{0,\alpha}\nabla_{p_{0,\alpha}}\right],$$
(noting that $\mathcal{P}L_{0}A=0$ for any $A$) and then we expand in the second term.  As shown in the SI (Sec.IV), we rigorously find that, for $d \rightarrow \infty$, all terms in the expansion are subleading. Thus we can replace the random forces $F_\alpha^\dagger(t)$ evolving with projected dynamics with the much simpler counterpart, $\tilde{F}_{0,\alpha}(t)$, evolving with $L_0$. Moreover, since $L_0$ does not contain $p_{0,\alpha}$, the gradient with respect to $p_{0,\alpha}$ of the force-force correlation in Eq.\ref{eq:3} vanishes.  
By replacing $p_{0,\alpha}(t)$ with $m\dot{u}_{0,\alpha}(t)$ in Eq.\ref{eq:3} we then obtain the Langevin equation
\begin{equation}\label{eq:4}
m\Ddot{u}_{0,\alpha}(t)=\tilde{F}_{0,\alpha}(t)-\frac{1}{k_{b}T}\int_{0}^{t} d\tau {\mathcal M}(\tau) \dot{u}_{0,\alpha}(t-\tau).
\end{equation}
Since the direction $\alpha$ is arbitrary, by isotropy, Eq.\ref{eq:4} is valid for any direction and thus can be seen as one component of a vector equation. To proceed further, we will first focus our analysis on a direction $\alpha$ which is fixed and uncorrelated from the 
interparticle directions  $\hat{\mathbf{R}}_{ij}(0)$.
In the large $d$ limit, the kernel ${\mathcal M}(\tau)=\langle \tilde{F}_{0,\alpha}(0)\tilde{F}_{0,\alpha}(\tau)\rangle_{0}$ reads
\begin{equation}\label{eq:kernel}
\left< \sum_{j,k}\hat{R}_{0j,\alpha}(0)\hat{R}_{0k,\alpha}(0)v'(\tilde{R}_{0j}(\tau))v'(\tilde{R}_{0k}(0))\right>_0,
\end{equation}
with the distance $\tilde{R}_{0j}(\tau)$ written as 
\begin{equation}\label{eq:X}
    \tilde{R}_{0j}(\tau)=R_{0j}(0)+\tilde y_{0j}(\tau)+\Delta_u(\tau)/2R_{0j}(0) 
\end{equation} 
where 
$\tilde y_{0j}(\tau)=\hat{\mathbf{R}}_{0j}(0) \cdot (\tilde{\bm{u}}_{0,\alpha}^{\perp}(\tau)-\tilde{\bm{u}}_j(\tau))$ and 
$\Delta_u(\tau)= (\tilde{\bm{u}}_{0,\alpha}^{\perp}(\tau)-\tilde{\bm{u}}_j(\tau))^2$. 
Eq.\ref{eq:X} is correct to order $1/d$, which is all that is required to evaluate the interaction potential $v(r)=\overline{v}(d(r/\ell-1))$. 
Note that the second term in the right-hand side of Eq. \ref{eq:X} is fluctuating in magnitude $\mathcal{O}(d^{-1})$, whereas the last term  concentrates on its average 
with sub-leading $\mathcal{O}(d^{-3/2})$ fluctuations (Sec.I of SI). 
The sum over off-diagonal, $j\neq k$,  contributions to the memory kernel in Eq.\ref{eq:kernel} can be neglected since it contains terms with random uncorrelated signs which, as shown in the SI (Sec.X), give a subleading  contribution \footnote{Note that this statement wouldn't hold if in the memory kernel the unblocked distance $R_{0j}(t)$ in stead of $\tilde{R}_{0j}(t)$ appeared}.
The final steps to obtain the DMFT expression of the memory kernel are demonstrating that for the diagonal contributions in Eq.\ref{eq:kernel}: (i) one can replace the restricted average $\langle \cdot \rangle_0$ by the full average $\langle \cdot \rangle$ and (ii) one can replace $\tilde{R}_{0j}(t)$ with $R_{0j}(t)$. We show in the SI (Sec.X) that this is indeed the case, i.e. the differences between these replacements and the original expressions are subleading in the large $d$ limit\footnote{It is important to stress that this is true {\it only for the diagonal terms} in the memory kernel, namely those which correlate the same pairs of particles at different times}. In conclusion the final expression of the memory kernel is
\begin{equation}\label{eq:mk}
{\mathcal M}(\tau)= \frac 1 d \sum_{j}\langle v'(R_{0j}(\tau))v'(R_{0j}(0))\rangle . 
\end{equation}
Note that the same reasoning also proves that  the fluctuating force $\tilde{F}_{0,\alpha}$ is a Gaussian random function as $\tilde{F}_{0,\alpha}$ sums up a large number $\mathcal{O}(d)$ of independent terms. In fact, by analyzing the higher-order averages of $\tilde{F}_{0,\alpha}(t)$ one can show that the leading contribution is obtained by pairing the particle indices in distinct couples. This effectively leads to Wick's theorem and Gaussian moments in $d\rightarrow\infty$. \\
\begin{figure}
\includegraphics[width=0.9\linewidth]{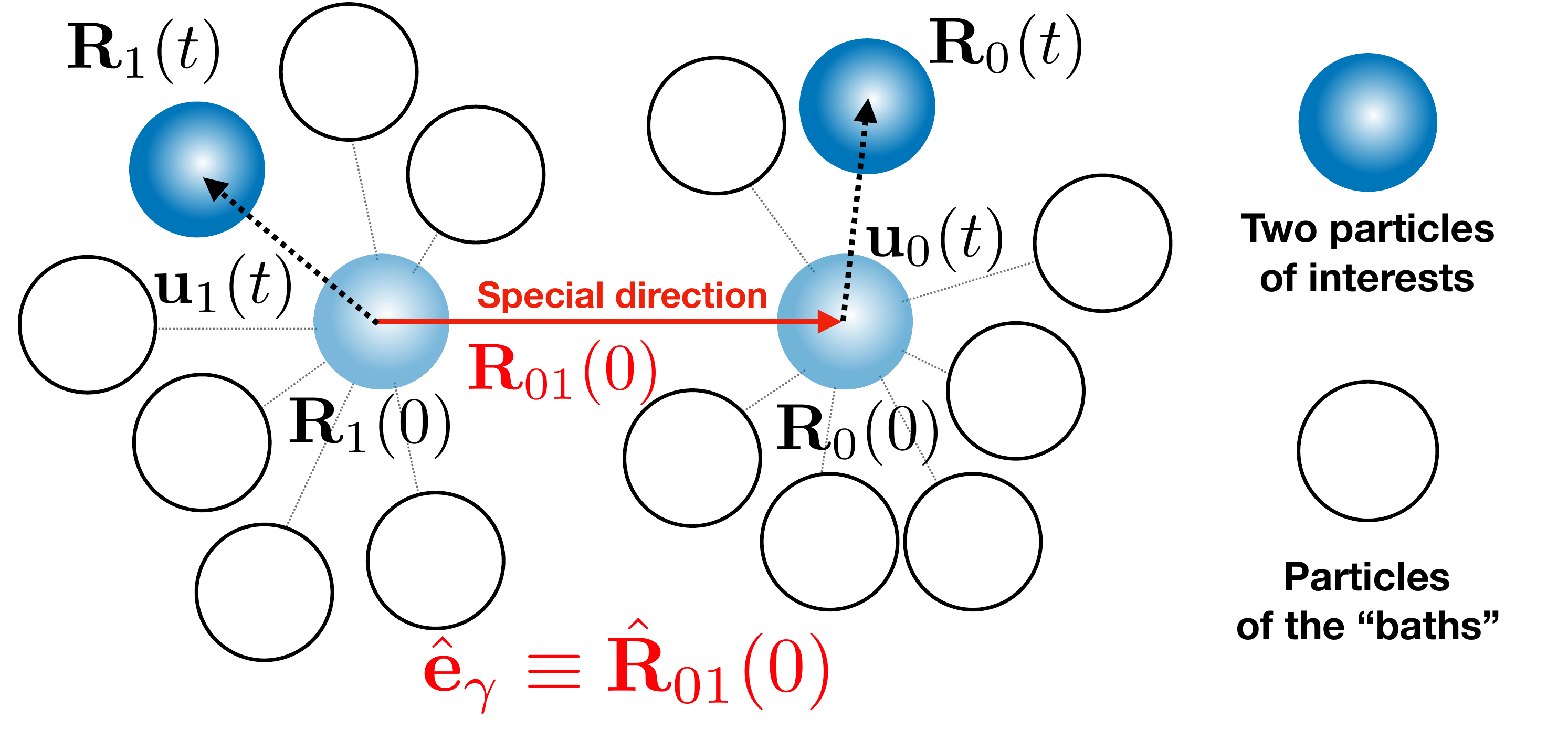}
\caption{Illustration of two particle dynamics and interactions which manifest in $d= \infty$.}
\label{fig:1}
\end{figure}

In order to obtain the full, closed DMFT equations, we must determine the equations for the {\em relative} displacement between a pair of particles for this is needed to compute the memory kernel (\ref{eq:mk}). Without loss of generality let's focus on particle $0$ and particle $1$ among the $j$-s in Eq.\ref{eq:mk}. Similarly to Eq. \ref{eq:X}, one can decompose $R_{01}(\tau)$ as
\begin{equation}\label{eq:R}
 R_{01}(\tau)=R_{01}(0)+w_{01}(\tau)+\Delta_w(\tau)/2R_{01}(0),
\end{equation} 
where
$w_{01}(\tau)=\mathbf{\hat R}_{01}(0) \cdot ({\bm{u}}_{0}(\tau)-{\bm{u}}_1(\tau))$ and 
$\Delta_w(\tau)= ({\bm{u}}_{0}(\tau)-{\bm{u}}_1(\tau))^2$. 
The squared displacements of particle $0$ and $1$ are self-averaging quantities for $d\rightarrow \infty$, i.e. they do not depend on the initial condition nor on the particle index. As a consequence, they are equal and can be obtained by the Langevin equation (\ref{eq:4}) leading to
\begin{equation}\label{eq:8}
    \Delta_w(\tau)\equiv2\Delta(t)=2d\langle (u_{0,\alpha}(t))^2 \rangle.
\end{equation}
The relative displacement of particles $0$ and $1$ along the direction $\hat{\bf R}_{01}(0)$, i.e. $w_{01}(\tau)$, is instead a fluctuating variable. This direction is special compared to $\alpha$ in that the force $-\bm{\nabla}v(R_{01}(\tau))$ acting on particle $0$ from particle $1$ is aligned with the initial condition (see Eq.\ref{eq:2} and remember that the initial conditions play the same role as quenched disorder). Thus, along direction $\hat{\bm{R}}_{01}(0)$, the contribution to the force on particle $0$ 
coming from the $(0,1)$ interaction is not small and needs to be treated explicitly. The interactions with all particles other than particle $1$ play the same role as before, as the direction $\hat{\bf R}_{01}$ is random for these particles thus leading to the same Gaussian random force and memory kernel of Eq.\ref{eq:4}, see Fig.\ref{fig:1}.  
The same reasoning extends to particle $1$. These considerations motivate the definition of a projector $\mathcal{P}_2$ in analogy with $\mathcal{P}$, for the dynamics where the motions of particle $0$ and $1$ are blocked along the special direction $\hat{\bm{e}}_\gamma\equiv\hat{\bm{R}}_{01}(0)$. Following Deutch and Oppenheim\cite{DO} (details in SI), we rigorously derive 
the equations for $u_{0,\gamma}\hat{=}\hat{\bm{e}}_\gamma\cdot \bm{u}_0$ and $u_{1,\gamma}\hat{=}\hat{\bm{e}}_\gamma\cdot \bm{u}_1$ which read
\begin{subequations}
\begin{multline}\label{Eq5}
m\Ddot{u}_{0,\gamma}(t) =-v'(R_{01}(0)+ w_{01}(t)+\Delta(t)/R_{01}(0))+ \tilde{F}_{0,\gamma}(t) \\
-\beta\int_{0}^{t} {\mathcal M}(\tau) \dot{u}_{0,\gamma}(t-\tau), 
\end{multline}
\begin{multline}\label{Eq6}
m\Ddot{u}_{1,\gamma}(t) =v'(R_{01}(0)+ w_{01}(t)+\Delta(t)/R_{01}(0))+ \tilde{F}_{1,\gamma}(t) \\
-\beta\int_{0}^{t} {\mathcal M}(\tau)\dot{u}_{1,\gamma}(t-\tau),
\end{multline}
\end{subequations}
where the random forces $\tilde{F}_{0,\gamma}$ and $\tilde{F}_{1,\gamma}$ are independent and Gaussian with covariance equal to the kernel ${\mathcal M}(\tau)$.  
By taking the difference between (\ref{Eq5}) and (\ref{Eq6}) one obtains a closed stochastic equation for $w_{01}(t)={u}_{0,\gamma}(t)-{u}_{1,\gamma}(t)$:
\begin{eqnarray}\label{Eq7}
  m\Ddot{w}_{01}(t) &=& -2v'(R_{01}(0)+ w_{01}(t)+\Delta(t)/2R_{01}(0)) 
\nonumber   \\ &&
+ \tilde{F}^{01}_{w}(t)
-\beta\int_{0}^{t} {\mathcal M}(\tau)\dot{w}_{01}(t-\tau),
\end{eqnarray}
where the random force $\tilde{F}^{01}_{w}(t)$ is Gaussian and with covariance $2\mathcal{M}(\tau)$\footnote{We emphasize that the covariance $\mathcal{M}(\tau)$ can be calculated using the statistics of the interparticle distances, Eq. (\ref{eq:mk}). However, it is the covariance of the forces evolving with unperturbed dynamics, $\tilde{F}(t)$, rather than with the full dynamics, $F(t)$.}.

We now have a full set of self-consistent equations. In fact, given $\Delta(t)$ and ${\mathcal M}(t)$, the stochastic Eqs. \ref{eq:4} and \ref{Eq7} for  $u_{0,\alpha}(t)$ and $w_{0j}(t)$ are fully determined. The stochastic processes associated with these equations allow us to obtain $\Delta(t)$ and ${\mathcal M}(t)$ as averages over $u_{0,\alpha}(t)$, $w_{01}(t)$ and the initial interparticle distances, see Eqs.\ref{eq:mk} and \ref{eq:8},  respectively. This closes the self-consistent loop. The DMFT equations therefore consist in the set of equations (\ref{eq:4}, \ref{eq:mk}, \ref{eq:R}, \ref{eq:8}, and \ref{Eq7}) which govern the evolution of the particle displacement and the distance between two particles. They can be simplified further, as done in  \cite{kurchan2016statics,agoritsas2018out} and detailed in SI (Sec.XII). 

The derivation based on projection operators is more involved for Brownian dynamics because of the presence of the noise which is a stochastic process independent of the dynamics of the system and thus equivalent to a set of external time-dependent forces acting on all the particles. It necessitates a generalization of the Liouville operator, which is now time dependent and which contains an additional term originating from adopting the Ito convention ~\cite{Ito}. After taking care of 
this and a few additional technical aspects, the derivation proceeds along the same lines as those in the Newtonian case and leads to the very same equations plus the usual additional friction and noise terms of Brownian motion (see Sec.VII and Sec.VIII of SI for details). 

Lastly, the cavity method \cite{mezard1987spin} also allows us to obtain the very same DMFT equations but follows an alternative, more explicit route. In particular, one writes down the full dynamical equations of motion in the absence of the cavity degree of freedom $u_{0,\alpha}(t)$, and treats the additional terms due to $u_{0,\alpha}(t)$ as a perturbation. At zeroth order the cavity degree of freedom only evolves due to the force $\tilde{F}_{0,\alpha}(t)$. In the large $d$ limit, one only needs to consider the linear order correction in perturbation theory since all other terms are subleading. Again, this is very similar to the derivation of the Langevin equation for a system coupled to a bath \cite{zwanzig1973nonlinear}. 
The force term $\tilde{F}_{0,\alpha}(t)$ is corrected to linear order since the dynamics of all the other particles is perturbed by $u_{0,\alpha}$. By taking into account this perturbatively linear correction one obtains the memory kernel Eq.\ref{eq:4}. From there the derivation follows the one we have sketched here. We refer to the SI (Sec.IX -- XI) for details. Note that one advantage of the cavity method compared to the previous derivations is that it allows to directly obtain DMFT equations valid also for non-equilibrium dynamics \cite{agoritsas2018out}. 

All three versions of the derivation rely upon the same insight: the identification of a variable that on the one hand allows us to describe the tagged particle dynamics but on the other hand perturbs the dynamics of the remaining degrees of freedom to sub-leading order of magnitude in $d$.  This leads to the possibility of neglecting higher-order corrections when $d\rightarrow \infty$. We note that had we chosen as our variable of interest ${\bf{u}}_{0}(t)$, namely the full displacement vector of the tagged particle, instead of $u_{0,\alpha}(t)$, then the perturbation theory would have lead to a series in which successive terms are of increasing, rather than decreasing, power of dimension (see the explicit discussion in Sec.IV of the SI). Thus, the DMFT equations found above can only be obtained if a component of the tagged particle coordinate is used. As a consequence, the pioneering work of Ref.\cite{agoritsas2019out} can only be considered as an approximate treatment. This fact is detailed in the Sec.IV and IX of SI. As a final note of caution, we stress that our whole derivation assumes time scales that do not diverge with $d$. This is a fundamental ingredient in all the scalings we used.

{\it Applications} -- Our simple approach to liquid-state DMFT opens up the possibility of the controlled description of the $d \rightarrow \infty$ dynamics for many systems, ranging from the dynamics of active particles to the single particle dynamics in random environment, \textit{e.g.} the random Lorentz gas ~\cite{lorentz1905mouvement}
The latter problem was recently analyzed starting from a $d$-dimensional system and then added one additional dimension rendering the system $d+1$ dimensional, see Biroli \textit{et al.}~\cite{BiroliL} for details. This approach was inspired by an earlier analysis of the spherical perceptron~\cite{minsky2017perceptrons} by Agoritsas \textit{et al.}~\cite{agoritsas2018out}. As outlined in Sec. XIII of the SI, our present approach offers an alternative derivation of the random Lorentz gas in the large dimensional limit. 
It relies upon recognizing one component of the moving particle's position as the tagged degree of freedom. Importantly, the present approach is more transparent and controlled in that it allows one to estimate the magnitude of terms neglected in the analysis. Finally, in Sec.XIV of the SI we sketch perhaps the most important use of the approach outline here, namely the extension to a cluster DMFT, which will be developed and analyzed in future work.

{\it Conclusions} -- 
In this paper we outline an interconnected set of direct and physically transparent routes to obtaining the exact dynamics of a liquid interacting via short ranged forces in $d\rightarrow\infty$. The unifying feature of these approaches is the identification of the tagged particle displacement along a {\em single} component of space as the ``cavity" variable whose influence on all other degrees of freedom is controllably small. Along with a dimensional scaling analysis, the use of this variable allows us to connect the cavity and projection operator methods of statistical mechanics, unify the behavior of Newtonian and Brownian fluids, and find facile solutions to the exact closed dynamics of venerable models of slow dynamics, such as that of the Lorentz gas in $d\rightarrow\infty$.  Future work will be focused in two directions.  First, our approach should provide a simple route to the full dynamics of other interesting liquid state problems in the high dimensional limit.  One such example is the behavior of active hard spheres, where the $d\rightarrow\infty$ steady-state properties have recently been explicated~\cite{Fred}.  Perhaps more ambitiously, we plan to take inspiration from the treatment of correlated electronic problems, where, for example, the exact behavior of local properties of the Hubbard model in $d\rightarrow\infty$ can be extended systematically to lower dimensions via a ``cluster" DMFT approach.  The scaling approach presented here enables the formulation of such cluster approaches for classical fluids, providing a promising path towards the grand challenge goal of a theory that can quantitatively treat glassy dynamics in low space dimensions.

{\it Acknowledgements} --We would like to thank E. Agoritsas, T. Arnoulx De Pirey, A. Manacorda, F. Van Wijland, and F. Zamponi for useful discussions. This work received funding from the Simons Collaboration “Cracking the glass problem” via 454935 (G. Biroli and C. Liu) and 454951 (D. R. Reichman and C. Liu). G. Szamel gratefully acknowledges the support
of NSF Grant No.~CHE 1800282.

\bibliography{article.bib}

\end{document}